\newif\ifANONYMOUS
\ANONYMOUStrue
\ANONYMOUSfalse

\newif\ifARXIV
\ARXIVtrue
\ARXIVfalse

\documentclass[conference]{IEEEtran}

\newif\ifDEBUG
\DEBUGtrue

\newif\ifAPPENDIX
\APPENDIXtrue
\APPENDIXfalse

\IEEEoverridecommandlockouts
\usepackage{cite}
\usepackage{graphicx}
\usepackage{textcomp}
\usepackage[frozencache,cachedir=.]{minted}
\usepackage{listings}
\usepackage{array}
\usepackage{orcidlink}
\newcolumntype{C}[1]{>{\centering\arraybackslash}p{#1}}

\def\BibTeX{{\rm B\kern-.05em{\sc i\kern-.025em b}\kern-.08em
    T\kern-.1667em\lower.7ex\hbox{E}\kern-.125emX}}

\usepackage{algpseudocode}
\usepackage[normalem]{ulem} 
\usepackage{xspace}
\usepackage{multirow} 
\usepackage{balance} 
\usepackage{fancyvrb} 
\usepackage{caption}
\usepackage{booktabs} 

\usepackage{enumitem}
\setlist[itemize]{leftmargin=*,noitemsep,topsep=0pt}
\setlist[enumerate]{leftmargin=*}

\usepackage{etoolbox}
\makeatletter
\patchcmd{\@makecaption}
	{\scshape}
	{}
	{}
	{}
\makeatletter
\patchcmd{\@makecaption}
	{\\}
	{.\ }
	{}
	{}
\makeatother
\def\tablename{Table}

\newcommand{\eg}{\textit{e.g.,}\xspace}
\newcommand{\etal}{\textit{et al.}\xspace}

\usepackage{mdframed}
\mdfsetup{skipabove=0.5\topskip,skipbelow=0.5\topskip,align=center}

\usepackage{amsthm}
\newtheorem{thm}{Theorem}

\DeclareMathSymbol{\mlq}{\mathord}{operators}{``}
\DeclareMathSymbol{\mrq}{\mathord}{operators}{`'}

\newif\ifSAVESPACE
\SAVESPACEfalse

\ifSAVESPACE

\else

\fi

\usepackage{todonotes}
\usepackage{soul}
\usepackage{fontawesome}
\usepackage{ragged2e}

\ifDEBUG
    \newcommand{\AH}[1]{\todo[color=cyan,inline]{AH:#1}}
    \newcommand{\AM}[1]{\todo[color=red,inline]{Machiry:#1}}
    \newcommand{\JD}[1]{\todo[color=yellow,inline]{JD:#1}}
    \newcommand{\TL}[1]{\todo[color=green,inline]{SA:#1}}
    \newcommand{\PA}[1]{\todo[color=orange,inline]{PA:#1}}
    
    \newcommand{\KR}[1]{\todo[color=yellow,inline]{Kyle:#1}}
    \newcommand{\LS}[1]{\todo[color=green,inline]{LS:#1}}
    \newcommand{\HP}[1]{\todo[color=green,inline]{HP:#1}}
    \newcommand{\PP}[1]{\todo[color=lime,inline]{PP: #1}}

\else
    \newcommand{\AH}[1]{}
    \newcommand{\AM}[1]{}
    \newcommand{\JD}[1]{}
    \newcommand{\TL}[1]{}
    \newcommand{\PA}[1]{}
    \newcommand{\KR}[1]{}
    \newcommand{\LS}[1]{}
    \newcommand{\HP}[1]{}
    \newcommand{\PP}[1]{}

\fi

\newcommand{\definition}[1]{
\begin{tcolorbox} [width=1.0\linewidth, colback=blue!07!white, top=1pt, bottom=1pt, left=2pt, right=2pt]
#1
\end{tcolorbox}
}

\newcommand{\observation}[1]{
\begin{tcolorbox} [width=1.0\linewidth, colback=pink!07!white, top=1pt, bottom=1pt, left=2pt, right=2pt]
#1
\end{tcolorbox}
}

\usepackage{hyperref}

\usepackage{cleveref}
\crefformat{section}{\S#2#1#3}
\crefname{figure}{Figure}{Figures}
\crefname{table}{Table}{Tables}
\crefname{theorem}{Theorem}{Theorems}
\crefname{thm}{Theorem}{Theorems}
\crefname{lemma}{Lemma}{Lemmata}
\crefname{equation}{Eqt.}{Eqts.}
\crefformat{Grammar}{Grammar #1}
\crefname{appendix}{Appendix}{Appendices}
\crefname{listing}{Listing}{Listings}

\newcommand{\myparagraph}[1]{\paragraph{#1}}
\renewcommand{\myparagraph}[1]{\vspace{0.25em} \noindent \underline{\textit{#1:}}}

\usepackage{url}

\usepackage{tcolorbox}

\makeatletter
\newcommand{\linebreakand}{%
  \end{@IEEEauthorhalign}
  \hfill\mbox{}\par
  \mbox{}\hfill\begin{@IEEEauthorhalign}
}
\makeatother

\usepackage{pifont}

 \usepackage{pifont}

\begin{document}

\title{A Research Agenda to Democratize Unit Proof Memory Safety Verification}
\title{A Research Agenda for Unit Proof Verification}
\title{A Unit Proof Engineering Framework for Memory Safety Verification}
\title{Unit Proof Engineering for Memory Safety Verification: A Research Agenda}
\title{Enabling Unit Proof Engineering for Memory Safety Verification: A Research Agenda}
\title{Enabling Unit Proofing for Software Verification: A Research Agenda}
\title{A Unit Proofing Framework for Software Implementation Verification: A Research Agenda}
\title{Enabling Unit Proofing for Code-level Verification}
\title{A Unit Proofing Framework for Code-level Verification: A Research Agenda}
\title{A Unit Proofing Framework for Code-level Verification: A Research Agenda}


\ifANONYMOUS
\author{
{\rm Anonymous author(s)}
}
\else

\author{
\IEEEauthorblockN{Paschal C. Amusuo\orcidlink{0000-0003-1001-525X}}
\IEEEauthorblockA{\textit{Purdue University}}
\and
\IEEEauthorblockN{Parth V. Patil\orcidlink{0009-0005-7337-1114}}
\IEEEauthorblockA{\textit{Purdue University}}
\and
\IEEEauthorblockN{Owen Cochell\orcidlink{0009-0003-5027-7710}}
\IEEEauthorblockA{\textit{Michigan State University}}
\and
\IEEEauthorblockN{Taylor Le Lievre\orcidlink{0009-0007-4816-9875}}
\IEEEauthorblockA{\textit{Purdue University}}
\and
\IEEEauthorblockN{James C. Davis\orcidlink{0000-0003-2495-686X}}
\IEEEauthorblockA{\textit{Purdue University}}

}

\fi

\maketitle

\begin{abstract}

Formal verification provides mathematical guarantees that a software is correct.
Design-level verification tools ensure software specifications are correct, but they do not expose defects in actual implementations. 
For this purpose, engineers use code-level tools.
However, such tools struggle to scale to large software.
The process of ``Unit Proofing'' mitigates this by decomposing the software and verifying each unit independently.
We examined AWS's use of unit proofing and observed that
  current approaches are manual
  and
  prone to faults that mask severe defects.
We propose a research agenda for a unit proofing framework, both methods and tools, to support software engineers in applying unit proofing effectively and efficiently.
This will enable engineers to discover code-level defects early.

\end{abstract}

\begin{IEEEkeywords}
Vision, Formal methods, Empirical software eng.
\end{IEEEkeywords}
\vspace{-0.1cm}

\vspace{-0.4cm}
\section{Introduction}
\vspace{-0.1cm}

Software underlies many critical digital and cyber-physical systems. 
Code-level defects, such as arithmetic errors and memory corruption vulnerabilities, compromise system safety and security and cause significant losses~\cite{anandayuvaraj_fail_2024}. 
For example, a null-pointer bug in CrowdStrike Falcon in 2024 impacted global transportation systems and caused over \$5 billion in damages~\cite{shweta_sharma_counting_nodate}. 
Traditional engineering practices, such as testing~\cite{amusuo_systematically_2023, mallissery_demystify_2023}, program analysis~\cite{pistoia_survey_2007}, and runtime sandboxing\cite{amusuo_ztd_java_2024}, can detect or mitigate such defects.
However, these methods validate software only approximately
and cannot guarantee correct behavior on all inputs. 
In contrast, formal verification~\cite{hoare_axiomatic_1969} proves software correctness, guaranteeing no code-level defects. 
This raises an important (and perennial) question~\cite{woodcock_formal_2009}: how to make formal verification more cost-effective for software engineers?

Engineering tools and processes help reduce the cost of applying a technology.
For formal methods, there are code-level verification tools, \eg the C and Rust bounded model checkers (CBMC\cite{kroening_cbmc_2014}, Kani~\cite{kani_getting_nodate}).
To improve both engineering and tool scalability,
  engineers decompose the software and verify individual functions within it via \textit{unit proofs}~\cite{moy_testing_2013}. 
These unit proofs model the function's interaction with other software components and verify the function using this model.
We refer to the process of developing, using, and maintaining unit proofs as \textit{\ul{unit proofing}}.
Organizations such as AWS~\cite{chong_code-level_2020} and ARM~\cite{wu_verifying_2024} have adopted unit proofing to ensure software correctness and memory safety. 

Since code-level verification depends on the unit proofs, they must be correct.
However, there is limited guidance and tool support for constructing unit proofs. 
Existing guidelines are \textit{ad hoc} and require expertise~\cite{bahig_formal_2017}. 
There is little publicly-available tool support, leading to costly manual development of unit proofs.
In~\cref{sec:unit-proof-challenges}, we assessed 11 unit proofs used to verify a widely-used operating system.
Although 5 of them correctly caught severe defects, 6 had faults that obscured others.

\textit{We believe that empirically-based guidelines and enhanced tool support will make unit proofing easier and enable early discovery of code-level defects.}
We propose a research agenda for a principled end-to-end unit proofing framework that can support software engineers in each stage of the unit proofing process. 
The proposed framework comprises of methods and tools and will help software engineers decompose a software into optimal verifiable units, develop correct unit proofs for each unit, validate and repair faulty unit proofs, and ensure unit proofs are updated as the software evolves.
We identify eight open questions that will need to be addressed, and share our future plans to this end.
If the proposed unit proofing framework helps software engineers verify their software code during development, then it will disrupt existing defect-finding practices such as static analysis, unit testing, and fuzzing.



\vspace{-0.3cm}
\begin{figure}[h]
    \centering
    \includegraphics[width=\columnwidth]{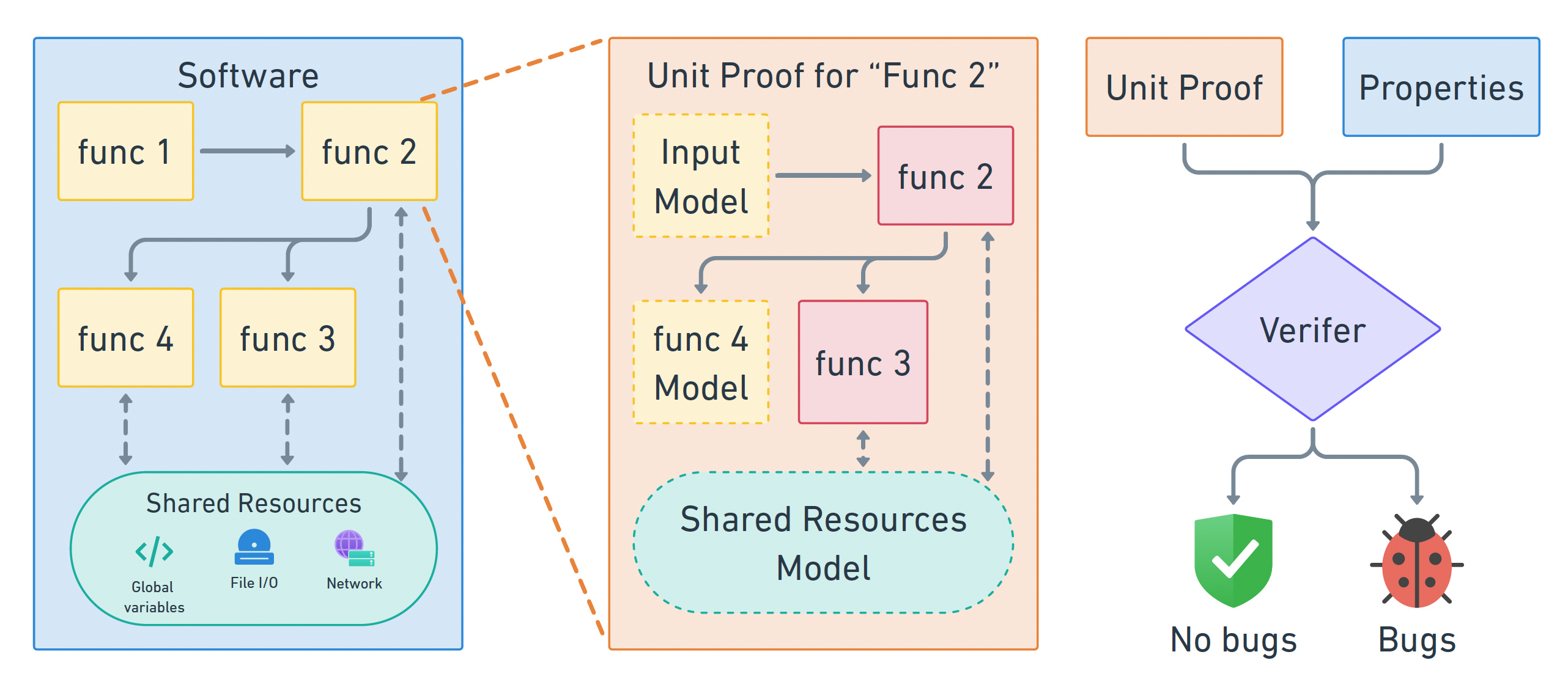}
    \vspace{-0.1cm}
    \caption{
    Overview of Unit Proofing.
    To verify function 2, an engineer 
      models its input, shared resources, and other functions. 
    A verifier can then check it in isolation.
    }
    \label{fig:unit-proof-overview}
    \vspace{-0.2cm}
\end{figure}

\vspace{-0.15cm}
\section{Code-level Verification and Unit Proofing}

This section introduces code-level verification (CLV) and unit proofing as a process for applying CLV on large software.


\vspace{-0.05cm}
\subsection{Formally Verifying Software Implementations}




\begin{listing}
  \centering
  \captionsetup{font=small}
  \caption{
   Verifying function implementations with a unit proof. Developers can specify functional properties (blue) and provide a model for a called function. Tools can introduce program-agnostic properties (red).
   Verification tools check that these properties hold.
  }
  \vspace{-0.1cm}
  \label{listing:function-verification}
  \begin{tcolorbox} [width=\linewidth, colback=white!30!white, top=1pt, 
  bottom=1pt, left=2pt, right=2pt]
\begin{minted}[
    fontsize=\scriptsize,
    linenos,
    gobble=2, % Remove unnecessary indentation -- the line numbers make this clear enough
    xleftmargin=0.5cm, % Otherwise we start in the left margin...
    escapeinside=||, % Allows you to use LaTeX commands inside code
    style=colorful,   % You can change this to another style like 'monokai', 'borland', etc.
    breaklines        % Break long lines
]{c}

int MODEL_get_max_len() { return rand(); }

char* get_ipv6_str(char *ipv6_str, uint8 *buffer) {
    |\textcolor{blue}{assume(valid\_pointer(ipv6\_str));}|
    struct in6_addr ipv6_addr;
    memcpy(&ipv6_addr, buffer, 16);
    |\textcolor{red}{assert(buffer != NULL);}|
    |\textcolor{red}{assert(sizeof(buffer) >= 16);}|
    int max_len = MODEL_get_max_len();
    ... // logic to convert buffer to IPv6 string
    |\textcolor{blue}{assert(valid\_ipv6(ipv6\_str));}|
}

\end{minted}

\end{tcolorbox}
\vspace{-0.2cm}
\end{listing}

Many tools verify software implementations using a variety of formal methods~\cite{wenzel_isabelle_2008, lattuada_verus_2023, kroening_cbmc_2014}.
This paper focuses on tools that use Bounded Model Checking (BMC)~\cite{clarke_bounded_2001} as they require the least human effort~\cite{silva_case_2012}, can automatically verify properties like memory safety, and are being adopted by major software organizations~\cite{chong_code-level_2020, wu_verifying_2024}.
We refer the reader to other works for comparisons of such tools~\cite{beckert_reasoning_2014, silva_case_2012}.

Bounded Model Checkers (BMCs) verify software correctness by ensuring all program states satisfy specific properties within a bound (\eg up to loop iteration count or recursion depth)~\cite{clarke_bounded_2001}.
As~\cref{listing:function-verification} shows, properties may be defined by developers (blue) or generated (red). BMC tools translate the software and properties to formal logic and use constraint solvers to verify the properties in all program states~\cite{martin_brain_cbmc_nodate}.

BMC has been applied
  in both industry (\eg AWS libraries~\cite{chong_code-level_2020}, ARM confidential computing firmware~\cite{wu_verifying_2024}) and open-source projects (\eg Firecracker, Hifitime, S2N-Quic~\cite{rahul_kumar_how_2023}). Beyer~\etal~\cite{beyer_software_2017} report that BMC tools find significantly more bugs in less time compared to testing and fuzzing tools.
  However, BMC struggles with state space explosions when applied to large software systems.

\subsection{Unit Proofing for Enabling Code-level Verification}
\label{subsec:unit-proofing}



To apply BMC tools in practice, several papers have described similar engineering processes~\cite{chong_code-level_2020, wu_verifying_2024}. 
One example is AWS's verification of embedded software and data structure libraries~\cite{chong_code-level_2020}.
In their process, engineers develop \textit{unit proofs} and use them to verify individual functions in the software. 
We refer to this process as \textit{``Unit Proofing''}.

Unit proofing comprise four stages. 
(1) Software is first decomposed into manageable units. 
(2) Unit proofs are then created to configure verification and model the unit's environment, including input, resources, and undefined functions (\cref{fig:unit-proof-overview}). 
(3) Initial proofs may include misconfigurations, inaccurate input or function models, causing prolonged verification, low coverage, or false positives and negatives. 
These faults are investigated and the proofs repaired.
(4) Finally, as software evolves, proofs must be updated to remain aligned with their respective units. 
Manually completing these steps, as reported by Chong~\etal~\cite{chong_code-level_2020}, is costly and error-prone.


Unit proofing shares similarities with \textit{unit testing} and \textit{compositional verification}.
Like \textit{unit tests}, AWS's unit proofs are developed together with the software, stored in the same repository, and executed in continuous integration workflows~\cite{chong_code-level_2020}.
However, unit tests check correctness on specific inputs, while unit proofs verify correctness across all possible inputs within the model~\cite{noauthor_cbmc_nodate}.
Compared with \textit{compositional verification}, unit proofing is an engineering process that enables compositional bounded model checking. 
There are works that automate compositional bounded model checking on real software~\cite{cho_blitz_2013, beckert_modular_2020} and abstract software models~\cite{bensalem_compositional_2010, cofer_compositional_2012}.
We focus on unit proofing due to its adoption by major software companies, including AWS and ARM.

\section{Exploratory Study of Unit Proofing}
\label{sec:unit-proof-challenges}


Many questions arise when considering any engineering process, such as ``Does it work?'' and ``What does it cost?''~\cite{kalu_reflecting_2023}. 
\textbf{While there are empirical evaluations of code-level tools on software benchmarks~\cite{beyer_software_2017}, no studies address their effectiveness on real software or within the unit proofing process.}
As a first step, we ask:

\noindent \textit{RQ: Do engineers' typical unit proofs expose targeted defects?}

We evaluated this question using FreeRTOS unit proofs and reported vulnerabilities (CVEs). FreeRTOS was chosen due to its widespread use and available unit proofs~\cite{noauthor_freertosfreertostestcbmc_nodate}. CVEs were used as they typically reflect critical memory safety vulnerabilities, which the unit proofs intend to address~\cite{noauthor_freertosfreertostestcbmc_nodate}.

\subsection{Methodology}
\textit{Dataset:} We collected all 21 CVEs affecting FreeRTOS from the National Vuln. Database (NVD) as of March 2024. We selected those affecting memory safety with sufficient details for reproduction, yielding 11 CVEs. For each, we identified the affected function and the associated unit proofs. 

\textit{Vulnerability Recreation:} Following the vulnerability description, we removed any input validations introduced in the function that fixed the vulnerability. 
Attempts to build and evaluate versions that predate the vulnerability fix failed, mostly due to unavailable or incompatible dependencies.


\textit{Unit Proof Assessment:} We executed the unit proofs using CBMC v5.95.1 and assessed whether the vulnerability was reported. If not, we repaired the unit proofs as needed. 

\subsection{Results}

Only 5/11 recreated vulnerabilities were exposed by FreeRTOS unit proofs (\cref{tab:freertos-flaws}).
With changes to the unit proofs, all vulnerabilities were exposed and one new high-severity vulnerability found (CVE-2024-38373). 
Unit proofs are effective, but unit proof engineers can err. 

\begin{table}
    \centering
    \caption{Assessing unit proofs on known vulnerabilities. Only 5 of the 11 vulnerabilities were exposed by the existing FreeRTOS unit proofs.
    \textit{UP}: Unit Proofing.
    }
    \begin{tabular}{C{2.1cm}C{0.7cm}C{2.5cm}C{2cm}}
    \toprule
         \textbf{Recreated Vuln} &  \textbf{Exposed}  & \textbf{Reason} & \textbf{UP Stage} \\
         \midrule
         CVE-2018-16523 &  No   & Insuff. properties & Design \\
         CVE-2018-16524 &  No   & Function Modeling & Development  \\
         CVE-2018-16525 &  No   & Out-of-date Proofs & Maintenance  \\
         CVE-2018-16527 &  No   & Insuff. Input Model & Development \\
         CVE-2018-16600 &  No   & Insuff. Input Model & Development   \\
         CVE-2018-16603 &  No   & Insuff. Input Model  & Development  \\
         \midrule
         CVE-2021-31571 &  Yes  & --- & --- \\
         CVE-2018-16526 &  Yes  & --- & --- \\
         CVE-2018-16599 &  Yes  & --- & --- \\
         CVE-2018-16601 &  Yes  & --- & --- \\
         CVE-2018-16602 &  Yes  & --- & --- \\
         \bottomrule
    \end{tabular}
    \label{tab:freertos-flaws}
\end{table}





\subsection{Observations on Unit Proofing Failures}

There were four reasons for the unit proofing failures.

\subsubsection{Insufficient Verification Properties (CVE-2018-16523)}

BMC tools can verify program-agnostic properties, but require the properties to be enabled. The unit proof for CVE-2018-16523, a division-by-zero defect, missed the issue because the division-by-zero property was disabled~\footnote{From CBMC v6, this property is enabled by default.}. The rationale behind disabling specific correctness checks remains unclear.


\subsubsection{Insufficient Input Modeling (\eg CVE-2018-16527)}

Unit proofs must model inputs broadly enough to cover all call sites yet precisely enough to avoid false positives. 
CVE-2018-16527 only occurs when the header of the input packet is truncated. 
The unit proof for the affected function incorrectly assumed the input packet always contained the header.


\subsubsection{Insufficient Function Modeling (CVE-2018-16524)}
\label{subsubsec:challenges-func-model}

To simplify verification, unit proofs use function models to replace specific called functions, retaining essential behaviors and properties. In CVE-2018-16524, an out-of-bound pointer computed in the verified function was only accessed in a replaced function. The replaced function's model failed to check the pointer's validity, leaving the vulnerability undetected.

\subsubsection{Out-of-date Unit Proofs (CVE-2018-16525)}

The vulnerable function's signature was changed but the new signature was not propagated to the unit proof. As CBMC supported making calls to undefined functions and only reported coverage on the reachable lines of code, the updated function was no longer verified.
When we fixed this unit proof, we exposed and reported a new vulnerability~\footnote{In response to our disclosure, CBMC's default behavior for undefined functions was changed to report errors instead of warnings. See \url{https://github.com/diffblue/cbmc/pull/8292}.
}.

\observation{
\textbf{Summary of Findings:} The default unit proofs only exposed 5/11 vulnerabilities. When fixed, all 11 CVEs were exposed. The gaps were caused by insufficient properties, insufficient modelling, and out-of-date proofs.
}

\begin{figure}
    \centering
    \includegraphics[width=0.8\columnwidth]{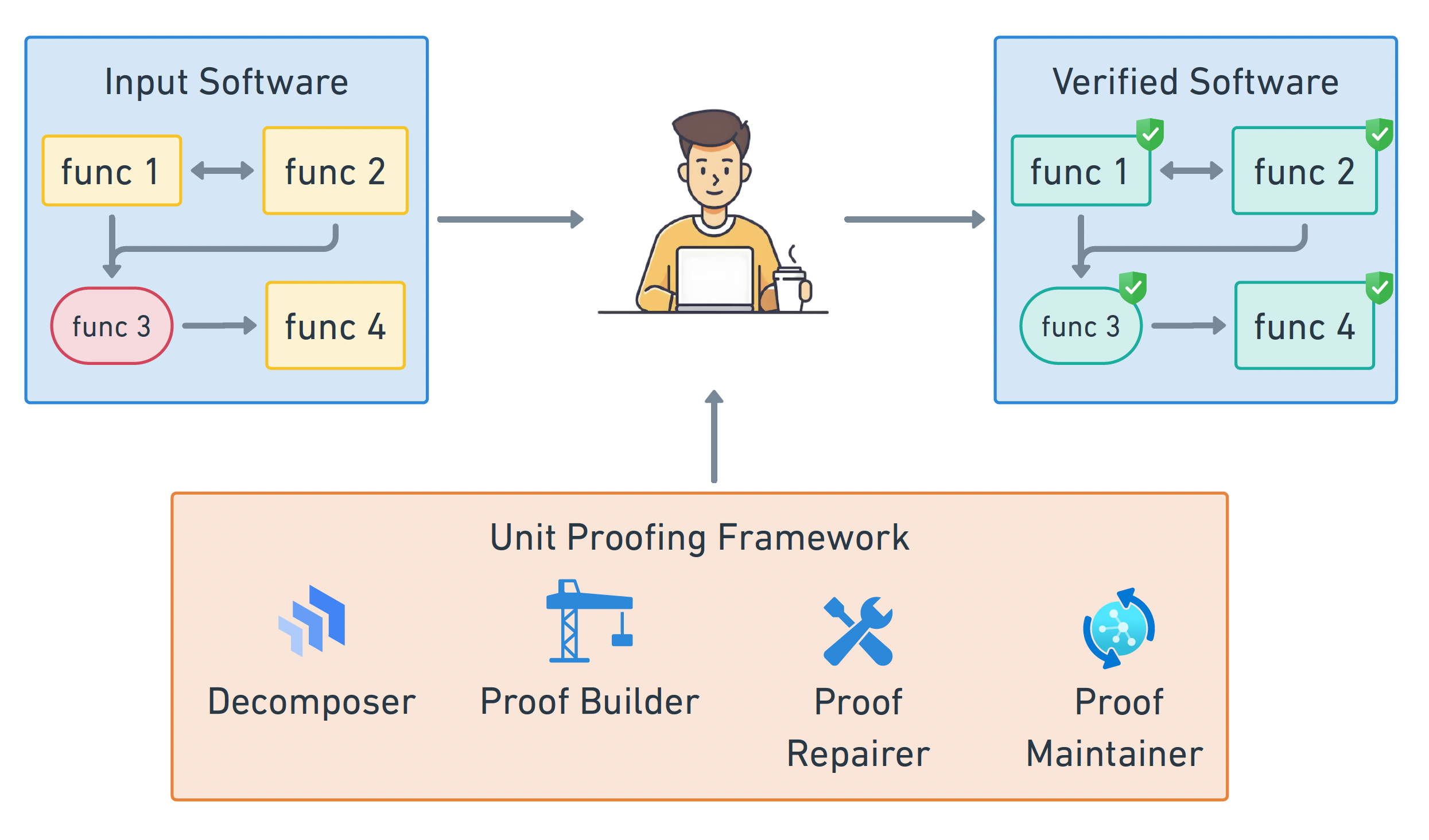}
    \caption{
    An end-to-end agenda for unit proof engineering.
    Software engineers use a set of tools and intelligent agents to verify the memory safety of applications.
    }
    \label{fig:unit-proof-agenda}
    \vspace{-0.3cm}
\end{figure}



\section{Research Agenda: A Unit Proofing Framework}

This section proposes a unit proofing framework, comprising four novel tools, to help software engineers develop and maintain unit proofs for accurate code-level verification. 
We first outline the goals of each proofing stage and identify open research questions. 
Then, we present the framework's design (\cref{fig:unit-proof-agenda}) and concrete research plans for implementation. 
Once completed, the tools will be integrated into standard software engineering workflows, facilitating early detection of code-level defects.

\subsection{Unit Proofing Stages and Open Questions}


Here, we discuss the unit proofing stages introduced in \cref{subsec:unit-proofing} and identify \textbf{O}pen \textbf{Q}uestions (OQs) within them.

\myparagraph{Software Decomposition}
The goal here is to minimize unit proofing effort but ensure each unit is solvable.

\definition{
\textbf{OQ1:} How can we estimate a function's verification time and UP effort?\\
\textbf{OQ2:} How should a software be decomposed optimally? 
}

Unit proofing involves decomposing software into function units and verifying each one. 
The decomposition strategy impacts the number of unit proofs, the complexity of their models, and verification tractability. 
Smaller units are solvable but require creating more unit proofs to cover the software. 
Additionally, the effort to create a unit proof depend on the complexity of the models it requires.
Units with complex input structures or that access multiple shared resources may require more effort to create unit proof models.
Hence, an optimal decomposition should minimize unit proofing efforts but ensure all units are solvable.

\myparagraph{Unit Proof Design and Development}
The goal here is to learn how to develop correct unit proofing models for functions with complex interactions with other software components.


\definition{
\textbf{OQ3:} How can we generate correct unit proof models? 
}

Unit proofs require models of a function's inputs, shared resources, and replaced functions. 
The complexity of these models depends on the complexity of interactions within the software.
The accuracy of the models affects verification properties such as coverage, soundness and completeness. 
Under-constrained models are easy to develop but lead to false positives~\cite{evgeny_novikov_klever__nodate}.
Conversely, over-constrained models hinder bug detection (\cref{sec:unit-proof-challenges}). 
The modeling complexity is increased in stateful software where shared state variables can be influenced by the function's callers and other functions in the software.

\myparagraph{Unit Proof Validation and Repair}
The goal here is to learn how to investigate and repair faults within unit proofs.

\definition{
\textbf{OQ4:} How do unit proof faults prevent defect exposure? \\
\textbf{OQ5:} How can we repair faulty unit proofs?
}

Faulty unit proof models or configurations can produce low coverage, false positives or prevent defect exposure.
Fixing these faults requires understanding the different kinds of faults, how they manifest and how they affect verification.
Additionally, fixing unit proof faults still rely on the developer expertise and experience as existing unit proofing guidelines~\cite{noauthor_writing_nodate} do not provide specific interventions for fixing faults.

\myparagraph{Unit Proof Evolution and Maintenance}
The goal here is to automatically update unit proofs as the software changes.

\definition{
\textbf{OQ7:} How do software changes affect unit proofs? \\
\textbf{OQ8:} What is the minimum unit proof change required to reverify a changed software?
}

Software evolves, and~\cref{sec:unit-proof-challenges} shows that failing to update unit proofs leads to incorrect verification. Since unit proofs verify interacting functions, changes in one proof can invalidate others.
Frequent software changes can make it impractical to regenerate or validate unit proofs after every change~\cite{chong_code-level_2020}.
Understanding how software changes will affect unit proofs, and repairing misaligned unit proofs, will ensure unit proofs remain consistent with the software.




\subsection{The Unit Proofing Framework Design}


Here, we present concrete research plans for actualizing each component of the proposed framework.

\subsubsection{Software Decomposer}
This tool will employ novel algorithms to minimize unit count while ensuring verification completes within a set time.
This requires new methods to estimate model complexity and verification time. 
Building on research linking program features to specification size and effort~\cite{staples_formal_2013, matichuk_empirical_2015}, we will analyze unit proofs to understand how program features influence complexity and duration. 
These insights will guide a cost estimator design that evaluates function features to estimate complexity and verification time and a decomposer algorithm to iteratively replace the most expensive functions with models. 
We will also explore a bottom-up approach using graph decomposition algorithms~\cite{zhang_using_2010, karande_bcd_2018} to partition call graphs, balancing verification time and modeling complexity to minimize overall costs.

\subsubsection{Proof Builder}
\label{subsubsec:proof-builder}


A unit proof requires sound and complete models for accurate verification. In stateful systems or those using shared resources, the resource model also depends on prior execution sequences. Traditional learning-based assume-guarantee reasoning~\cite{cobleigh_learning_2003, giannakopoulou_assumption_2002} refines assumptions iteratively using counterexamples, but applying this to functions with multiple interactions is costly due to the need for repeated model checking~\cite{cobleigh_breaking_2008}.
To address this, we propose to use a property-violation guided learning approach, starting with generic proof models and refining the models only until all reachable code is covered and reported violations are resolved. Refinements will be informed by interventions from the \textit{proof repairer}~(\cref{subsubsec:proof-repairer}). 
The learned models will then be propagated as assertions to the respective components, with violations indicating defects. 
This approach will also help indicate specific execution paths that trigger a defect.


\subsubsection{Proof Repairer}
\label{subsubsec:proof-repairer}

The proof repairer will detect faults in unit proofs and intelligently suggest fixes. 
To design the repairer, we will develop a taxonomy of unit proof faults and interventions by empirically studying existing and new unit proofs, assessing their coverage, violation reports, and bug detection abilities, and identifying necessary interventions for different faults.
Investigating and fixing proof faults requires a deep understanding of program semantics.
Hence, using the developed taxonomy, the repairer will detect faults by combining the fast but imprecise program analysis techniques the code reasoning abilities of large language models (LLMs)~\cite{fang_large_2024, nam_using_2024}. 
We will explore two hybrid approaches: one using search-based repairs with LLMs as a fallback for precision, and another enhancing repair tools with insights from LLMs. Both approaches will be evaluated for effectiveness.

\subsubsection{Proof Updater}
The proposed proof updater will keep unit proofs synchronized with software changes. When the software changes, the updater will identify affected unit proofs, assess their validity, and apply necessary updates. 
An empirical study will examine how software changes impact proofs, what invalidates them, and how they are fixed. 
To validate updates, the updater will reverify affected functions, comparing results with prior data using metrics like coverage and violation reports.
Initially, the \textit{proof repairer} (\cref{subsubsec:proof-repairer}) will be used to resolve invalidated proofs. 
In future iterations, we plan to develop an LLM agent that, informed by empirical insights, can propose specific updates to the proof that realigns it with the changed software.


\section{Future Plans}

\myparagraph{Ongoing Work}
We have two active efforts towards the proof builder and proof repairer.

First, we are evaluating the effectiveness and cost of the property-violation guided learning approach for the proof builder by manually developing unit proofs for functions in four selected software, identifying necessary interventions for coverage gaps and property violations, and assessing defect classes exposed by the unit proofs. Insights from this study have been used to refine unit proof development guidelines.

Second, we are leveraging program analysis and LLMs to automate proof building and repair. The proof repairer uses fault insights and interventions from the first study to fix generic unit proofs. We plan to assess the accuracy of the developed proofs against manually developed ones and their performance in developing proofs for new software.

\myparagraph{Longer-Term} 
Our research agenda will take many people many years. 
We invite the software engineering research community to join us.
Together, we can develop a world where every function is verified as it is being written.



\section{Conclusion}

In this paper, we present a research agenda for a unit proofing framework that will enable engineers verify their software implementations. The framework is motivated by our exploratory study of AWS's unit proofs. It comprises tools and methods to decompose a software, develop and repair unit proofs, and keep proofs updated as the software evolves.

\textit{Acknowledgement}: This project is being supported by a Rolls Royce research grant.

\textit{Data availability:}
  Our artifact contains the data for~\cref{sec:unit-proof-challenges}.
  See \url{https://doi.org/10.5281/zenodo.14660073}.



\clearpage

{\footnotesize \bibliographystyle{unsrt}
\bibliography{cbmc_project_edited}}


\vspace{12pt}

\end{document}